# Improved variational approach in semileptonic decay of *B* meson for the Cornell potential

S. Rahmani[1] and H. Hassanabadi*[1]

[1]*Physics Department, Shahrood University, Shahrood, Iran*

* Email: h.hassanabadi@shahroodut.ac.ir

**Abstract**

We obtain the wave function of mesonic systems interacting via a Cornell potential by applying improved variational approach. We study Isgur-Wise function and its parameters. Then we report our calculations on the parameters of semileptonic decay of *B* and $B_s$ mesons. $B \to D^{(*)} l \nu$ process is also investigated in this work.

## 1. Introduction

There are many of the problems in physics which cannot be solved exactly. Common methods dealt to approximately to solve these problems which are the perturbation theory, the variational method and also combinations of them. The variational calculation is constructed for quantum mechanical problems [1]. The success of the variational method depends on our consideration of wave function how which can be treating. Variational method can be readily applied in evaluating excited states. Variational method can be applied in many aspects and problems of physics i.e. analytical treatment of excited states for radial wave functions [2,3], problem of screening an external charge in strongly correlated metals [4], maximizing entropy to the problem of out of equilibrium physical systems [5], quantum field theory [6] and heavy quarkonium properties [7].

Cornell interaction (Coulomb plus linear) that is static, spherically symmetric interaction has a physically application in mesonic systems ie. charmonium and bottomonium [7-9]. Coulomb-like part is a short range that arises from exchanging a massless gluon between the quarks whereas linear part is a long range which responsible for confinement phenomena that linearly term is assumed to summarize all nonperturbative effects of the strong interactions. Coulombic interaction is known from perturbative quantum chromodynamics and the large distance interaction known from lattice QCD [10,11]. Improved variational approach was proposed to solve radial Schrödinger equation for the Cornell potential [3]. Variationally improved perturbation theory is also applied to the Cornell potential [9]. In fact the variationally improved perturbation theory method was proposed by Ref. [12] and later developed in the paper of Aitchison and Dudek for heavy quarkonia systems [9].

Among various models to investigate structure of mesons Isgur-Wise function (IWF for short) stands as a capable tool. Where the charm and bottom quarks are considered infinitely heavy, their strong interactions with light quarks and gluons acquire additional symmetries [13]. Relativistic treatment quark model [14], Bethe-Salpeter approach [15], QCD sum rules [16], are used to study of mesonic systems. One can be found study on the *B* meson decays from IWF approach in Refs. [17-23].

Next section we will review on our calculation of the variational method. In section (3) we show the IWF and parameters for bottom $B$ and charmed $D$ mesons and use them to investigate the semileptonic $B$ meson decay. In section (4) we discuss on our numerical results.

## 2. Wave function

We consider Hamiltonian with Cornell interaction

$$H = -\frac{\hbar^2}{2\mu}\nabla^2 + \alpha r + \frac{\beta}{r} + V_0 \tag{1}$$

which cannot be solved exactly. Hence to find eigenenergies and eigenfunctions we proceed to variational method. We take the initial wave function as Hydrogen-type

$$\psi_{n,l}(a,r) = N_{nor}(ar)^{l+1} e^{-ar} L_n^{2l+1}(ar) \tag{2}$$

where $L_n^{2l+1}(ar)$ are the generalized Laguerre functions and $a$, $N_{nor}$ are the variational parameter and normalization constant of the wave function, respectively. We get only ground state related to $n = 1$, $l = 0$. If we calculate the virial condition as

$$\bar{E} = \frac{<\psi(ar)|H|\psi(ar)>}{<\psi(ar)|\psi(ar)>} \tag{3}$$

then for ground state we receive at

$$\bar{E}_{1,0}(a) = \frac{5a^3 + 75\mu\alpha + 6a^2\mu\beta}{18a\mu} \tag{4}$$

Now minimizing Eq. (4) as low expectation value as possible,

$$\frac{\partial \bar{E}_{1,0}(a)}{\partial a} = 0 \tag{5}$$

we can get variational parameter for different heavy-light mesons. Having eigenvalues we can report masses of some heavy-light mesons as shown in Table 1.

## 3. $B$ and $B_s$ semileptonic Decays, Isgur-Wise Function

The differential decay rate for the semileptonic decay $B \to Dl\nu$ process is defined by [24,25]

$$\frac{d\Gamma(B \to D\ell\nu)}{d\omega} = |\bar{\eta}_{EW}|^2 \frac{G_F^2}{48\pi^3}|V_{cb}|^2 (m_B + m_D)^2 m_D^3 (\omega^2 - 1)^{\frac{3}{2}} \Im^2(\omega) \tag{6}$$

in the limit of very small lepton masses ($l = e$ or $\mu$) with $\omega$ is the dot product of the four-velocities of the $B$ and $D$ mesons. $|\bar{\eta}_{EW}|^2$ stands for electroweak corrections. $G_F$ is the Fermi coupling constant, and $m_B$ and $m_D$ are the masses of the $B$ and $D$ mesons, respectively. $\Im(\omega)$ is a single form factor that in the limit of infinite quark masses coincides with the Isgur-Wise function $\xi(\omega)$. The normalization of this form factor is known in the zero recoil limit, where the initial and final meson have the same velocity. In fact current conservation implies the normalization of the IWF. We use the expansion form of IWF near the zero recoil point parameterized as [26]

$$\xi(\omega) = 1 - \rho^2(\omega - 1) + C(\omega - 1)^2 + ... \tag{7}$$

where the so-called slope and curvature of IWF define as

$$\rho^2 = 4\pi\mu^2 \int_0^\infty r^4 |\psi_{1,0}(a,r)|^2 \, dr, \tag{8}$$

$$C = \frac{2}{3}\pi\mu^4 \int_0^\infty r^6 |\psi_{1,0}(a,r)|^2 \, dr \tag{9}$$

respectively. We have presented the obtained parameters in Table 2. Using of Eq. (7) in Eq. (6) and integrating it over the range

$$1 \leq \omega \leq \frac{m_B^2 + m_D^2}{2 m_B m_D} \tag{10}$$

we can report on $B$ meson semileptonic decay width. The same way is done to report $B_s$ meson semileptonic decay width. In the $m_{b,c} \Box \Lambda_{QCD}$ limit the six form factors correspond to semileptonic decay $B \to D^{(*)} l\nu$, are determined by a single universal IWF denoted by $\xi(\omega)$ [25]. Here $\omega$ is the dot product of the four-velocities of the $B$ and $D^{(*)}$ mesons. We define for the semileptonic decay $B \to D^{(*)} l\nu$ process [25]

$$\frac{d\Gamma(B \to D^{(*)} l\nu)}{d\omega} = \frac{G_F^2 |V_{cb}|^2 m_B^5}{48\pi^3} r^3 (1-r)^2 \sqrt{\omega^2 - 1}(\omega + 1)^2$$
$$\times \left[1 + \frac{4\omega}{\omega + 1} \frac{1 - 2r\omega + r^2}{(1-r)^2}\right] [\xi(\omega)]^2 \tag{11}$$

where $r = \frac{m_{D^{(*)}}}{m_B}$. We have calculated decay width, branching ratio and $|V_{cb}|$ for semileptonic $B$ decays in Table 3.

## 4. Results and discussion

In the present work we have considered the improved variational method to investigate $B \to Dl\nu$, $B_s \to D_s l\nu$ and $B \to D^{(*)} l\nu$ processes in the heavy quark limit. In this way we have obtained eigenfunctions and eigenvalues of the mesonic system in the nonrelativistic prescription with the choice of the Cornell potential.

As IWF is the overlapping of wave function of two hadrons it can be written as [27] $\xi(\omega) = \sqrt{\frac{2}{\omega + 1}} \langle \psi_D | \psi_B \rangle$. In this case the slope and curvature of IWF can be obtained as 0.24 and 0.18 respectively. In our numerical calculations we have taken the potential parameters as $\alpha = 1.383 \, GeV^2$ and $\beta = -0.293$. We have used the experimental value of $B$ meson mass to find $V_0$ as $V_0 = -5.688 GeV$. The masses of bottom and charmed $B$ and $D$ mesons are taken as $m_{\bar{B}} = 5.279\ GeV$, $m_D = 1.869\ GeV$, $m_{\bar{B}_s} = 5.366\ GeV$, $m_{D_s} = 1.968\ GeV$ and $m_{D^*} = 2.006\ GeV$ in the calculations [28]. We have also taken $\tau_B = 1.638 ps$ and $\tau_{B_s} = 1.512 ps$ [28]. We have compared our results in Table 2 with [20,29]. To extract $/V_{cb}/$ from $B \to D\ell\nu$ semileptonic decay we consider $\bar{\eta}_{EW} = 1.011$ [24] in Eq. (6). Thereby adopting the experimental value for $Br(B \to D\ell\nu)$ of Ref. [28] and our approach for IWF in the previous section we have reported $/V_{cb}/$ as 0.040 with uncertainty of 4.7‰ in comparison of available value of [28]. In the case of $B_s \to D_s l\nu$ we have used the reported value of $Br(B_s \to D_s\ell\nu) = 1.4$ [30] and we have gained $/V_{cb}/ = 0.039$. Some of reported values of branching ratio for the semileptonic decay $B \to D\ell\nu$ from different approaches are $2.31 \pm 0.09$ in the work of Bernlochner et al. [25], $2.34 \pm 0.03 \pm 0.13$ of the work of the BABAR Collaboration [31], $2.15 \pm 0.06 \pm 0.09$ reported by Palombo [32]. We have compared our results for $B$ semileptonic decays with Refs. [19,28,30,33] in Table 3.

As can be seen form Tables and mentioned references the agreement of the calculated quantities with existing experimental data and other theoretical values is satisfactory.

Table 1. Meson masses

| Mesons | Mass [this work] | Mass [28] |
|---|---|---|
| $D$ | 2.145 | 1.869 |
| $D_s$ | 1.980 | 1.968 |
| $B_s$ | 5.037 | 5.366 |

Table 2. Slope and curvature of IWF

| Mesons | $D$ | $B$ | $D_s$ | $B_s$ |
|---|---|---|---|---|
| $\rho^2$ [this work] | 0.73 | 0.87 | 1.03 | 1.30 |
| $\rho^2$ [others] | 0.68 [29] | 0.70 [29] | 1.07 [29] | 1.46 [20] |
| $C$ [this work] | 0.13 | 0.19 | 0.26 | 0.42 |
| $C$ [29] | 0.11 | 0.11 | 0.28 | 0.28 |

Table 3. Decay width, branching ratio and $|V_{cb}|$ for semileptonic $B$ decays

| Process | $B \to Dl\nu$ | $B_s \to D_s l\nu$ | $B \to D^{(*)}l\nu$ |
|---|---|---|---|
| $\Gamma$ [this work] | 1.35 | 0.96 | 4.04 |
| $\Gamma$ [others] | 1.413 [19] | - | 4.651 [19] |
| $Br$ [this work] | 2.21 | 1.46 | 6.62 |
| $Br$ [others] | 2.27 ± 0.11 [28] | 1.4 - 1.7 [30] | 6.5 ± 0.5 [33] |
| $|V_{cb}|$ [this work] | 0.040 | 0.039 | 0.039 |
| $|V_{cb}|$ [28] | 0.042 ± 0.001 | | |